\documentclass{Interspeech}
\usepackage{multirow}
\usepackage{makecell}
\usepackage{bbding}

\interspeechcameraready 

\title{Multi-Channel Sequence-to-Sequence Neural Diarization: Experimental Results for The MISP 2025 Challenge}

\author[affiliation={1}]{Ming}{Cheng}
\author[affiliation={1}]{Fei}{Su}
\author[affiliation={1}]{Cancan}{Li}
\author[affiliation={1,2}]{Juan}{Liu}
\author[affiliation={1,3}]{Ming}{Li}

\affiliation{School of Computer Science}{Wuhan University}{China}
\affiliation{School of Artificial Intelligence}{Wuhan University}{China}
\affiliation{Suzhou Municipal Key Laboratory of  Multimodal Intelligent Systems, Digital Innovation Research Center}{Duke Kunshan University}{China}
\email{ming.li369@dukekunshan.edu.cn}


\usepackage{comment}

\begin{document}

\maketitle

\begingroup
  \renewcommand\thefootnote{}
  \footnote{Corresponding Author: Ming Li}
  \addtocounter{footnote}{-1}
\endgroup

\begin{abstract}

This paper describes the speaker diarization system developed for the Multimodal Information-Based Speech Processing (MISP) 2025 Challenge. First, we utilize the Sequence-to-Sequence Neural Diarization (S2SND) framework to generate initial predictions using single-channel audio. Then, we extend the original S2SND framework to create a new version, Multi-Channel Sequence-to-Sequence Neural Diarization (MC-S2SND), which refines the initial results using multi-channel audio. The final system achieves a diarization error rate (DER) of 8.09\% on the evaluation set of the competition database, ranking first place in the speaker diarization task of the MISP 2025 Challenge.

\end{abstract}

\section{Introduction}

Speaker diarization refers to the process of identifying each speaker's utterance boundaries in conversational data, addressing the ``Who-Spoke-When'' problem~\cite{park2022review}. It is essential for various downstream speech-related tasks, such as multi-talker speech recognition~\cite{kanda2020joint}.

Many classical studies have been proposed for speaker diarization. Conventional approaches, often referred to as modularized methods, rely on independent modules to split the audio signal into short segments and cluster speaker identities based on speaker embedding similarities~\cite{wang2018speaker, lin2019lstm, landini2022bayesian}. However, these methods struggle with overlapped speech, as the clustering module inherently assumes that each audio segment should be speaker-homogeneous. 

End-to-End Neural Diarization (EEND) and its variants~\cite{fujita2019end_1, horiguchi2020end, horiguchi2022encoder} can predict the voice activities of multiple speakers through multi-label classification, demonstrating strong robustness to overlapped speech. Nevertheless, the Permutation-Invariant Training (PIT)~\cite{hershey2016deep} employed in EEND models tends to degrade performance as the number of speakers increases in long audio signals, which has yet to be fully resolved.

Target-Speaker Voice Activity Detection (TSVAD) approaches~\cite{medennikov2020target, wang2022similarity, cheng2023target} combine modularized methods with end-to-end neural networks. A typical TSVAD system first uses the modularized method to extract target-speaker embeddings as speaker enrollment. Then, it employs a back-end neural network to predict the voice activities of all speakers. TSVAD-based methods have demonstrated outstanding performance on popular benchmarks such as DIHARD-III~\cite{wang2021ustc} and VoxSRC21-23~\cite{wang2021dku, wang2022dku, cheng2023dku}. 

Recently, a novel Sequence-to-Sequence Neural Diarization (S2SND) framework~\cite{cheng2024sequence} has been proposed for speaker diarization. By utilizing an automatic speaker detection and representation technique, it eliminates the need for unsupervised clustering methods (e.g., K-Means~\cite{wang2018speaker}, SC~\cite{lin2019lstm}, AHC~\cite{sell2018diarization}) and the PIT~\cite{hershey2016deep} strategy. The proposed S2SND models significantly outperform previous state-of-the-art methods, achieving lower diarization error rates on the DIHARD-II~\cite{ryant2019second} and DIHARD-III~\cite{ryant2020third} evaluation sets.

\begin{figure}[t]
\centering
  \includegraphics[width=0.95\linewidth]{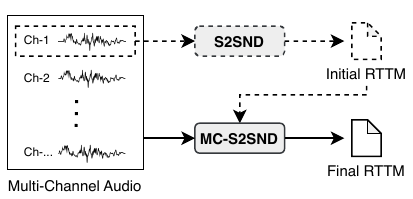}
  \caption{Overview of our developed system. The dashed parts represent the prior single-channel diarization method~\cite{cheng2024sequence}.}
  \label{fig:overview}
\end{figure}
  
In order to advance speaker diarization performance for the MISP 2025 Challenge~\cite{gao2025multimodalinformationbasedspeech}, we choose the promising S2SND framework~\cite{cheng2024sequence} as the foundation. However, the S2SND model is primarily designed for online inference using single-channel audio, which may not be ideal for offline inference using multi-channel audio. In this paper, we extend the S2SND framework to develop a new version tailored for offline multi-channel inference, named Multi-Channel Sequence-to-Sequence Neural Diarization (MC-S2SND). 

Fig.~\ref{fig:overview} illustrates the overview of our developed system. First, the original S2SND model processes the first-channel audio signal to obtain initial results. Next, the proposed MC-S2SND model leverages these initial results to extract speaker embeddings and then processes multi-channel audio to predict the outcomes. Experimental results demonstrate that the multi-channel system yields significant improvements over previous approaches, achieving a diarization error rate (DER) of 8.09\% on the evaluation set of the MISP 2025 Challenge.

\begin{figure*}[t]
\centering
  \includegraphics[width=\linewidth]{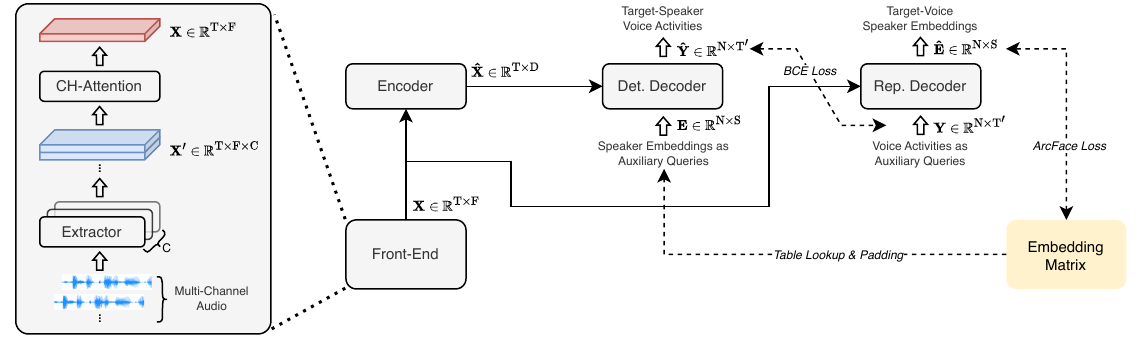}
  \caption{Multi-Channel Sequence-to-Sequence Neural Diarization (MC-S2SND) framework. \textit{Det.} and \textit{Rep.} denote the abbreviations of detection and representation, respectively.}
  \label{fig:framework}
\end{figure*}

\section{Methodology}

Fig.~\ref{fig:framework} demonstrates the proposed Multi-Channel Sequence-to-Sequence Neural Diarization (MC-S2SND) framework, which is partially modified from the original S2SND~\cite{cheng2024sequence} model. The details are described as follows.

\subsection{Architecture}
\label{sec:arch}

\subsubsection{Front-End}

The front-end module comprises the extractor and channel attention (CH-Attention) block. The extractor is based on a ResNet model~\cite{he2016deep} integrated with segmental statistical pooling (SSP)~\cite{wang2022similarity}. For the input audio with $C$ channels, each channel is converted into log Mel-filterbank energies and separately passed through the extractor. This step results in deep features denoted as $\mathbf{X}^{\prime} \in \mathbb{R}^\mathrm{T\times F\times C}$, where $T$ and $F$ represent the length and dimension of the extracted feature sequence. Subsequently, we implement a Transformer-based model~\cite{vaswani2017attention} combined with average pooling to form the channel-attention block. This step applies self-attention along the channel axis of $\mathbf{X}^{\prime}$ and averages the output across the different channels. The final feature sequence, which integrates multi-channel information, is represented as $\mathbf{X} \in \mathbb{R}^\mathrm{T\times F}$.

\subsubsection{Encoder}
 
The Conformer-based model~\cite{gulati2020conformer} with sinusoidal positional encodings~\cite{vaswani2017attention} is used as the encoder. Let $D$ denote the attention dimension in the encoder, with an additional linear layer that maps the input dimension from $F$ to $D$ (this layer is omitted for clarity in the plot). The encoder then processes the deep features $\mathbf{X} \in \mathbb{R}^\mathrm{T\times F}$ to produce $\mathbf{\hat{X}} \in \mathbb{R}^\mathrm{T\times D}$, which aims to model long-term dependencies within the frame-wise feature sequence.
 
\subsubsection{Decoder}

The structures of the detection (Det.) and representation (Rep.) decoders are detailed in the original S2SND work~\cite{cheng2024sequence}. Generally, these decoders take a set of auxiliary queries as reference information to output results for different speakers. Let $\mathbf{E} \in \mathbb{R}^\mathrm{N\times S}$ represent the speaker embeddings, where $N$ is the number of speakers and $S$ is the embedding dimension. The ground truth for their target voice activities is represented by a binary matrix $\mathbf{Y} \in \left\{0, 1\right\}^\mathrm{N\times T^{\prime}}$, where $y_{n, t^{\prime}}$ indicates whether the $n$-th speaker is speaking at time $t^{\prime}$. The detection decoder uses the encoder output $\mathbf{\hat{X}}$ as feature embeddings and speaker embeddings $\mathbf{E}$ as auxiliary queries to predict the target-speaker voice activities $\mathbf{\hat{Y}} \in \left\{0, 1\right\}^\mathrm{N\times T^{\prime}}$. On the other hand, the representation decoder uses the front-end output $\mathbf{X}$ as feature embeddings and the voice activities $\mathbf{Y}$ as auxiliary queries to extract the target-voice speaker embeddings $\mathbf{\hat{E}} \in \mathbb{R}^\mathrm{N\times S}$. The two decoders perform inverse tasks, predicting voice activities and extracting speaker embeddings simultaneously.

\subsection{Training Process}

The ground truth for $\mathbf{Y}$ can be obtained from the dataset annotations during training. However, $\mathbf{E}$ is not directly available because the embedding space must be learned by the neural network. Thus, we initialize a learnable embedding matrix $\mathbf{E}_\mathrm{all} \in \mathbb{R}^\mathrm{N_{all} \times S}$ to store the embedding vectors for all speakers in the training data, where $N_\mathrm{all}$ represents the total number of speakers and $S$ is the embedding dimension. Speaker labels are tokenized into $N_\mathrm{all}$-dimensional one-hot vectors. For example, the $n$-th speaker label is represented by a one-hot vector with all zeros except for the $n$-th position, which is set to 1. Given an input audio with speaker labels $\mathbf{S}_\mathrm{loc} \in \mathbb{R}^\mathrm{N_{loc} \times N_{all}}$, where $N_\mathrm{loc}$ is the number of locally existing speakers. The input speaker embeddings for the detection decoder are computed using $(\mathbf{S}_\mathrm{loc} \cdot \mathbf{E}_\mathrm{all}) \in \mathbb{R}^\mathrm{N_{loc} \times S}$, which is a simple table lookup operation through matrix multiplication. Also, a learnable non-speech embedding $\mathbf{e}_\mathrm{non}$ is initialized. To handle cases where $N_\mathrm{loc} \le N$, the absent input is padded with either $\mathbf{e}_\mathrm{non}$ or randomly selected embeddings from speakers not present in the current audio. The ground truth for these padded inputs is set to zeros. This padding strategy ensures that the input dimension of mini-batched training data remains consistent and allows the model to differentiate between valid (existing) and invalid (padded) speaker enrollment in the input audio. It is noteworthy that the masked speaker prediction technique used in the original S2SND model has been removed in the MC-S2SND model, as it is unnecessary for an offline-only model.

\subsubsection{Target-Speaker Voice Activity Detection}
 
After feeding $\mathbf{E}$ into the detection decoder, the output $\mathbf{\hat{Y}}$ is optimized by minimizing its Binary Cross-Entropy (BCE) loss with $\mathbf{Y}$, as defined by the following equation:

\begin{align}
\begin{split}
\mathcal{L}_{\mathrm{bce}} = - \frac{1}{N\times T^{\prime}} \sum_{n=1}^{N} \sum_{t^{\prime}=1}^{T^{\prime}} 
& \left[y_{n,t^{\prime}} \log(\hat{y}_{n,t^{\prime}}) + \right. \\ 
& \left. (1 - y_{n,t^{\prime}}) \log(1 - \hat{y}_{n,t^{\prime}}) \right],
\end{split}
\label{eq:bce}
\end{align}

\noindent where $\hat{y}_{n,t^{\prime}} = \mathbf{\hat{Y}}(n,t^{\prime})$ represents the predicted speaking probability of the $n$-th speaker at time $t^{\prime}$, and $y_{n,t^{\prime}} = \mathbf{Y}(n,t^{\prime})$ is the corresponding ground-truth label.

\subsubsection{Target-Voice Speaker Embedding Extraction}

After feeding $\mathbf{Y}$ into the representation decoder, the output $\mathbf{\hat{E}}$ is optimized by minimizing its ArcFace~\cite{deng2019arcface} loss with the embedding matrix $\mathbf{E}_\mathrm{all}$, as given by the following equation:

\begin{align}
\mathcal{L}_{\mathrm{arc}} = \frac{1}{N} \sum_{n=1}^{N} -\log\frac{e^{s\cos(\theta_{n}+m)}}{e^{s\cos(\theta_{n}+m)}+\sum_{i=1,i\neq S_{n}}^{N_{all}}e^{s\cos\theta_{i}}},
\label{eq:arcface} 
\end{align}

\noindent where $\theta_{n}$ is the angle between the $n$-th extracted speaker embedding $\mathbf{\hat{e}}_{n} \in \mathbf{\hat{E}}$ and its ground-truth embedding in $\mathbf{E}_\mathrm{all}$; $\theta_{i}$ is the angle between $\mathbf{\hat{e}}_{n}$ and the $i$-th speaker embedding in $\mathbf{E}_\mathrm{all}$. Let $S_{n}$ denote the index of the $n$-th given speaker label corresponding to $\mathbf{\hat{e}}_{n}$. The condition $i \neq S_{n}$ ensures that $\theta_{i}$ is computed only for negative pairs in this contrastive learning setup. The parameters $s$ and $m$ represent the re-scaling factor and additive angular margin penalty, respectively.

Finally, the total training loss is the sum of $\mathcal{L}_{\mathrm{bce}}$ in Eq.~\ref{eq:bce} and $\mathcal{L}_{\mathrm{arc}}$ in Eq.~\ref{eq:arcface}. This way, the input embedding space for speaker detection and the output embedding space for speaker representation can be jointly optimized within a single model.

\subsection{Inferring Process}
\label{sec:infer}

During inference, $\mathbf{Y}$ can be obtained from an initial diarization system. The following steps are then performed: First, the input multi-channel audio is processed through the front-end module and encoder to obtain the deep features $\mathbf{X}$ and $\mathbf{\hat{X}}$. Second, using the prepared $\mathbf{X}$ and $\mathbf{Y}$, the representation decoder extracts the speaker embeddings $\mathbf{\hat{E}}$ for enrollment. Third, the trained embedding matrix is no longer required. Using the prepared $\mathbf{\hat{X}}$ and $\mathbf{\hat{E}}$, the detection decoder predicts the corresponding voice activities, which form the final diarization result $\mathbf{\hat{Y}}$. Overall, the proposed framework integrates speaker embedding extraction and voice activity detection into a single model.

\section{Experimental Setup}

\subsection{Datasets}

For the simulated data, we combine the VoxCeleb2~\cite{Chung18b}, VoxBlink2~\cite{lin2024voxblink2}, KeSpeech~\cite{tang2021kespeech}, and 3D-Speaker~\cite{zheng20233d} datasets to create a large-scale corpus with 153,738 identities. Then, an on-the-fly simulation method can generate new training data, which is used in our previous works~\cite{cheng2023target,cheng2024sequence,cheng2024multi}. For the real data, the MISP-Meeting~\cite{chen2025misp} dataset is provided by the MISP 2025 Challenge, which includes 119 hours of training data, 3 hours of development data, and 3 hours of evaluation data, recorded using a headset microphone, microphone array, and panoramic camera. In this study, we only use the 8-channel audio data from the far-field microphone array and apply the NARA-WPE~\footnote{\url{https://github.com/fgnt/nara_wpe}} toolkit for dereverberation. The training set is used for model training, while the development set is reserved for model validation and hyperparameter tuning.

\subsection{Network Configurations}

As described in Sec.~\ref{sec:arch}, the main network architecture consists of the extractor, channel-attention, encoder, and decoder modules. The ResNet-152~\cite{he2016deep} model is used as the extractor, with residual blocks having widths (number of channels) of $\left\{64, 128, 256, 512\right\}$. The channel-attention module is based on a Transformer model~\cite{vaswani2017attention}, consisting of 2 blocks with 512-dimensional, 8-head attention and 1024-dimensional feedforward layers. The encoder is based on a Conformer~\cite{gulati2020conformer} model, utilizing convolutions with a kernel size 15. Both decoders are based on the Speaker-wise Decoder (SW-D) proposed in the original S2SND~\cite{cheng2024sequence}. All encoders and decoders have 6 blocks, and their settings of attention and feedforward layers are identical to those of the channel-attention module. The key difference between the S2SND and MC-S2SND models lies in the inclusion of the channel-attention module in the MC-S2SND model.

\subsection{Training Details}

All training audio is divided into 8-second blocks with a 2-second shift. After normalization with a mean of 0 and a standard deviation of 1, each audio block is converted into 80-dimensional log Mel-filterbank energies as the acoustic features with a frame length of 25 ms and a shift of 10 ms.

We directly adopt the default settings from the previous paper~\cite{cheng2024sequence}: the temporal resolution (duration per frame-level prediction) of the system output is set to 10 ms, and the speaker capacity $N$ is set to 30 to accommodate the maximum number of speakers in most cases. During training, the input speaker embeddings and voice activities are randomly shuffled to ensure the model is invariant to speaker order. As a result, the corresponding ground-truth labels must be reassigned according to the shuffled order. Additionally, data augmentation is performed using additive noise from Musan~\cite{snyder2015musan} and reverberation from RIRs~\cite{ko2017study}. The model is trained using 8 NVIDIA RTX-3090 GPUs with the AdamW~\cite{loshchilov2017decoupled} optimizer, Binary Cross-Entropy (BCE) loss, and ArcFace loss ($s=32, m=0.2$)~\cite{deng2019arcface}, as illustrated in Fig.~\ref{fig:framework}.

\subsubsection{Pretrained Extractor}

To enable the model to learn identity-related information in the audio effectively, we pretrain the extractor as a speaker verification model. Using the statistical pooling~\cite{snyder2018x} and 256-dimensional linear output layer, the extractor is trained with the ArcFace loss ($s=32, m=0.2$)~\cite{deng2019arcface} classifier on the VoxBlink2~\cite{lin2024voxblink2} dataset. Then, it undergoes large-margin fine-tuning (LMFT)~\cite{9414600} on the VoxCeleb2~\cite{Chung18b} dataset, achieving an equal error rate (EER) of 0.3403\% on the Vox-O~\cite{Nagrani17} trial.

\subsubsection{S2SND Model}

The S2SND model training follows the same procedure outlined in the paper~\cite{cheng2024sequence}. The process consists of several stages, which are briefly introduced as follows.

\begin{itemize}
\setlength{\itemsep}{0pt}
	\item Stage 1: The weights of the pretrained extractor are frozen, and the remaining parts of the model are trained solely on the simulated data with a learning rate of \textit{1e-4}.
	\item Stage 2: The weights of the pretrained extractor are unfrozen, and the entire S2SND model is trained using 50\% simulated data and 50\% real data.
	\item Stage 3: The learning rate is further decayed to \textit{1e-5} for the final fine-tuning.
\end{itemize}

\subsubsection{MC-S2SND Model}

Regarding network architecture, the MC-S2SND model differs from the original S2SND model only by adding a new channel attention module. Therefore, we begin training the MC-S2SND model using the weights from the previously trained S2SND model. Due to the lack of multi-channel simulated data, only real data is used for this model. The two-stage training strategy is described as follows:

\begin{itemize}
\setlength{\itemsep}{0pt}
	\item Stage 1: The weights initialized from the previous S2SND model are frozen, and the newly initialized channel attention module is added. The MC-S2SND model is trained with a learning rate of \textit{1e-4}.
	\item Stage 2: All weights of the MC-S2SND model are unfrozen and fine-tuned with a learning rate of \textit{1e-5}.
\end{itemize}

\subsection{Inferring Details}

The primary inference process has been introduced in Sec.~\ref{sec:infer}. However, due to limited GPU memory, all test audio must also be split, normalized, and converted into log Mel-filterbank energies in the same manner as the training data. For both S2SND and MC-S2SND models, two implementation tricks for blockwise processing are outlined as follows.

\subsubsection{Embedding Clustering}

At the step of speaker embedding extraction, each speaker will have multiple embedding vectors from different blocks. To maximize intra-speaker similarity and minimize inter-speaker similarity across all extracted embeddings, we perform K-Means clustering using the Scikit-Learn~\footnote{\url{https://scikit-learn.org/stable/}} toolkit. Specifically, the average embedding of each speaker is used as the initial centroid for the K-means algorithm. After clustering, the number of speakers remains unchanged, but some embedding vectors may be reassigned. The updated average embedding for each speaker is then used for voice activity detection.

\subsubsection{Score-level Fusion}

At the step of voice activity detection, the output from each block is stitched along the time axis. Since the block shift is shorter than the block length, there will be overlapping predictions between adjacent blocks. We average the overlapping predictions as a score-level fusion. The impacts of these techniques are evaluated in the experimental results.

\begin{table}[t]
	\centering
	\setlength{\tabcolsep}{4.5pt}
	\caption{Diarization error rates (DERs) of different models on the MISP-Meeting evaluation set.}
	\label{tab:exps}
	\begin{tabular}{llllr}
		\toprule
		\textbf{\#} &
		\multirow{2}{*}{\textbf{Method}} & 
		\multirow{2}{*}{\textbf{\makecell[l]{Embedding\\Clustering}}} &
		\multirow{2}{*}{\textbf{\makecell[l]{Block\\Shift}}} &	
		\multirow{2}{*}{\textbf{\makecell[l]{DER\\(\%)}}} \\
		\\
		\midrule
		1 & \multirow{4}{*}{S2SND} & $\times$     & 8s  & 13.79 \\
		2 &                       & $\checkmark$ & 8s  & 13.68 \\
		3 &                       & $\times$     & 2s  & 11.89 \\
		4 &                       & $\checkmark$ & 2s  & 11.48 \\

		\midrule
		5 & \multirow{4}{*}{MC-S2SND} & $\times$     & 8s  & 9.77 \\
		6 &                          & $\checkmark$ & 8s  & 9.42 \\
		7 &                          & $\times$     & 2s  & 8.94 \\
		8 &                          & $\checkmark$ & 2s  & 8.62 \\	
		
		\midrule
		9 & MC-S2SND ($+$\textit{adapt.}) & $\checkmark$ & 2s & 8.09 \\ 
	
		\bottomrule
	\end{tabular}
\end{table}

\begin{figure}[t]
\centering
  \includegraphics[width=0.8\linewidth]{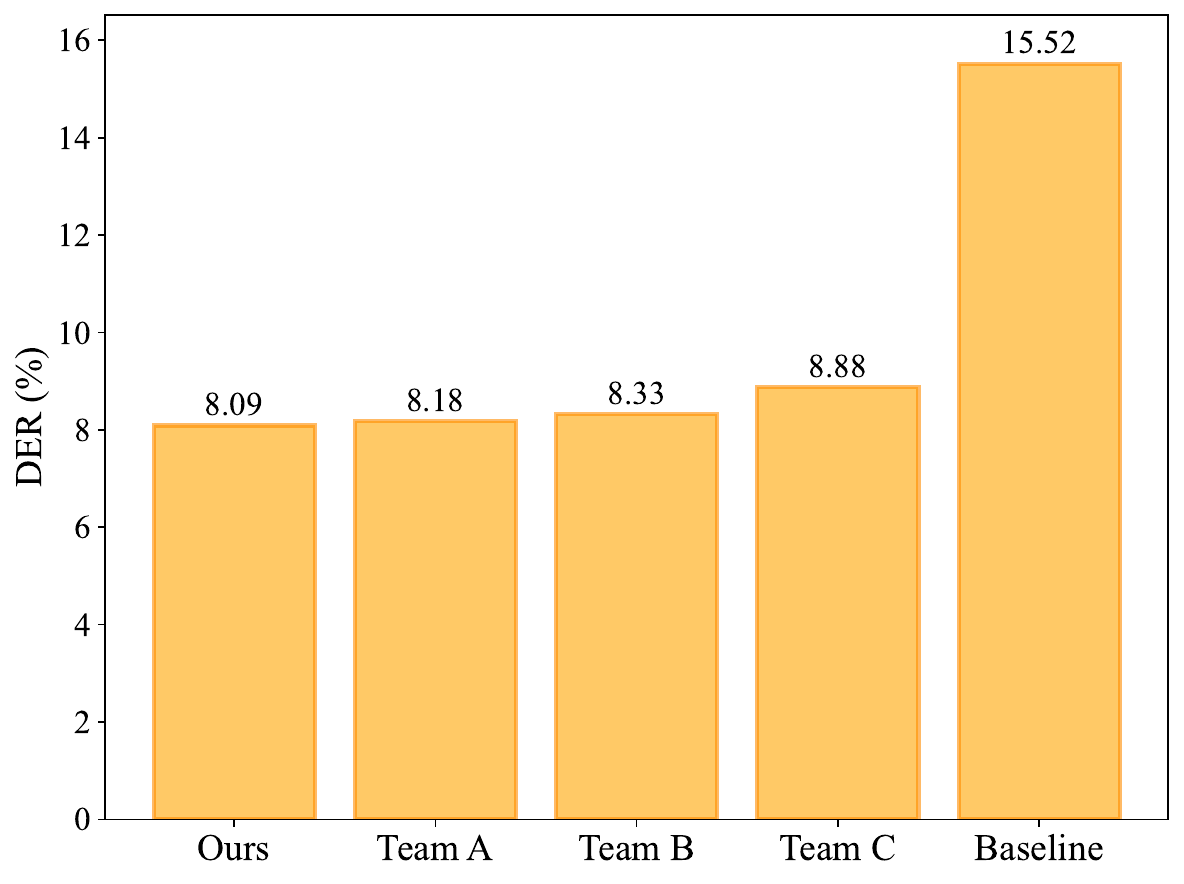}
  \caption{Leaderboard of the MISP 2025 Challenge (Task 1). The official baseline and top-ranked systems are plotted, where team names have been anonymized.}
  \label{fig:ders}
\end{figure}

\section{Results}

The evaluation metric is the diarization error rate (DER) without collar tolerance. Table~\ref{tab:exps} illustrates the performance of the systems we developed on the MISP-Meeting evaluation set. 

Systems \#1-4 represent the S2SND model with different inference settings. The ablation results show that the embedding clustering and dense score-level fusion strategies contribute to improved diarization performance. System \#4, which combines both strategies, achieves the best S2SND result with a diarization error rate (DER) of 11.48\%.

Systems \#5-8 represent the MC-S2SND model with different inference settings. The initial diarization result is provided by System \#4. Ablation experiments still demonstrate the effectiveness of embedding clustering and score-level fusion strategies. System \#8 achieves the best MC-S2SND result with a diarization error rate (DER) of 8.62\%.

The experimental results above show that our newly proposed MC-S2SND model significantly outperforms the original S2SND model in offline multi-channel inference. Finally, System \#9 further adapts System \#8 on a mixture of the training and development sets for several epochs. The resulting diarization error rate (DER) of 8.09\% demonstrates that deep neural networks can still greatly benefit from real high-quality data as much as possible.

\section{Conclusions}

This paper describes our proposed Multi-Channel Sequence-to-Sequence Neural Diarization (MC-S2SND) framework for the speaker diarization task in the MISP 2025 Challenge. Compared to the original S2SND method, the modified MC-S2SND model effectively processes multi-channel audio to enhance diarization performance in offline scenarios. As shown in Fig.~\ref{fig:ders}, the best result from our developed system ranks first place in Task 1 (speaker diarization) of the competition leaderboard.

\section{Acknowledgements}
This research is funded in part by the National Natural Science Foundation of China (62171207), Yangtze River Delta
Science and Technology Innovation Community Joint Research Project (2024CSJGG01100), Science and Technology
Program of Suzhou City(SYC2022051) and Guangdong Science and Technology Plan (2023A1111120012). Many thanks
for the computational resource provided by the Advanced Computing East China Sub-Center.

\bibliographystyle{IEEEtran}
\bibliography{refs}

\begin{thebibliography}{10}
\providecommand{\url}[1]{#1}
\csname url@samestyle\endcsname
\providecommand{\newblock}{\relax}
\providecommand{\bibinfo}[2]{#2}
\providecommand{\BIBentrySTDinterwordspacing}{\spaceskip=0pt\relax}
\providecommand{\BIBentryALTinterwordstretchfactor}{4}
\providecommand{\BIBentryALTinterwordspacing}{\spaceskip=\fontdimen2\font plus
\BIBentryALTinterwordstretchfactor\fontdimen3\font minus \fontdimen4\font\relax}
\providecommand{\BIBforeignlanguage}[2]{{%
\expandafter\ifx\csname l@#1\endcsname\relax
\typeout{** WARNING: IEEEtran.bst: No hyphenation pattern has been}%
\typeout{** loaded for the language `#1'. Using the pattern for}%
\typeout{** the default language instead.}%
\else
\language=\csname l@#1\endcsname
\fi
#2}}
\providecommand{\BIBdecl}{\relax}
\BIBdecl

\bibitem{park2022review}
T.~J. Park, N.~Kanda, D.~Dimitriadis, K.~J. Han, S.~Watanabe, and S.~Narayanan, ``A review of speaker diarization: Recent advances with deep learning,'' \emph{Computer Speech \& Language}, vol.~72, p. 101317, 2022.

\bibitem{kanda2020joint}
N.~Kanda, Y.~Gaur, X.~Wang, Z.~Meng, Z.~Chen, T.~Zhou, and T.~Yoshioka, ``Joint speaker counting, speech recognition, and speaker identification for overlapped speech of any number of speakers,'' in \emph{Proc. INTERSPEECH}, 2020, pp. 36--40.

\bibitem{wang2018speaker}
Q.~Wang, C.~Downey, L.~Wan, P.~A. Mansfield, and I.~L. Moreno, ``Speaker diarization with lstm,'' in \emph{Proc. ICASSP}, 2018, pp. 5239--5243.

\bibitem{lin2019lstm}
Q.~Lin, R.~Yin, M.~Li, H.~Bredin, and C.~Barras, ``Lstm based similarity measurement with spectral clustering for speaker diarization,'' in \emph{Proc. INTERSPEECH}, 2019, pp. 366--370.

\bibitem{landini2022bayesian}
F.~Landini, J.~Profant, M.~Diez, and L.~Burget, ``Bayesian hmm clustering of x-vector sequences (vbx) in speaker diarization: Theory, implementation and analysis on standard tasks,'' \emph{Computer Speech \& Language}, vol.~71, p. 101254, 2022.

\bibitem{fujita2019end_1}
Y.~Fujita, N.~Kanda, S.~Horiguchi, K.~Nagamatsu, and S.~Watanabe, ``End-to-end neural speaker diarization with permutation-free objectives,'' in \emph{Proc. INTERSPEECH}, 2019, pp. 4300--4304.

\bibitem{horiguchi2020end}
S.~Horiguchi, Y.~Fujita, S.~Watanabe, Y.~Xue, and K.~Nagamatsu, ``End-to-end speaker diarization for an unknown number of speakers with encoder-decoder based attractors,'' in \emph{Proc. INTERSPEECH}, 2020, pp. 269--273.

\bibitem{horiguchi2022encoder}
S.~Horiguchi, Y.~Fujita, S.~Watanabe, Y.~Xue, and P.~García, ``Encoder-decoder based attractors for end-to-end neural diarization,'' \emph{IEEE/ACM Transactions on Audio, Speech, and Language Processing}, vol.~30, pp. 1493--1507, 2022.

\bibitem{hershey2016deep}
J.~R. Hershey, Z.~Chen, J.~Le~Roux, and S.~Watanabe, ``Deep clustering: Discriminative embeddings for segmentation and separation,'' in \emph{Proc. ICASSP}, 2016, pp. 31--35.

\bibitem{medennikov2020target}
I.~Medennikov, M.~Korenevsky, T.~Prisyach, Y.~Khokhlov, M.~Korenevskaya, I.~Sorokin, T.~Timofeeva, A.~Mitrofanov, A.~Andrusenko, I.~Podluzhny, A.~Laptev, and A.~Romanenko, ``Target-speaker voice activity detection: A novel approach for multi-speaker diarization in a dinner party scenario,'' in \emph{Proc. INTERSPEECH}, 2020, pp. 274--278.

\bibitem{wang2022similarity}
W.~Wang, Q.~Lin, D.~Cai, and M.~Li, ``Similarity measurement of segment-level speaker embeddings in speaker diarization,'' \emph{IEEE/ACM Transactions on Audio, Speech, and Language Processing}, vol.~30, pp. 2645--2658, 2022.

\bibitem{cheng2023target}
M.~Cheng, W.~Wang, Y.~Zhang, X.~Qin, and M.~Li, ``Target-speaker voice activity detection via sequence-to-sequence prediction,'' in \emph{Proc. ICASSP}, 2023, pp. 1--5.

\bibitem{wang2021ustc}
Y.~Wang, M.~He, S.~Niu, L.~Sun, T.~Gao, X.~Fang, J.~Pan, J.~Du, and C.-H. Lee, ``Ustc-nelslip system description for dihard-iii challenge,'' \emph{arXiv preprint arXiv:2103.10661}, 2021.

\bibitem{wang2021dku}
W.~Wang, D.~Cai, Q.~Lin, L.~Yang, J.~Wang, J.~Wang, and M.~Li, ``The dku-dukeece-lenovo system for the diarization task of the 2021 voxceleb speaker recognition challenge,'' \emph{arXiv preprint arXiv:2109.02002}, 2021.

\bibitem{wang2022dku}
W.~Wang, X.~Qin, M.~Cheng, Y.~Zhang, K.~Wang, and M.~Li, ``The dku-dukeece diarization system for the voxceleb speaker recognition challenge 2022,'' \emph{arXiv preprint arXiv:2210.01677}, 2022.

\bibitem{cheng2023dku}
M.~Cheng, W.~Wang, X.~Qin, Y.~Lin, N.~Jiang, G.~Zhao, and M.~Li, ``The dku-msxf diarization system for the voxceleb speaker recognition challenge 2023,'' in \emph{Proc. NCMMSC}, 2024, pp. 330--337.

\bibitem{cheng2024sequence}
M.~Cheng, Y.~Lin, and M.~Li, ``Sequence-to-sequence neural diarization with automatic speaker detection and representation,'' \emph{arXiv preprint arXiv:2411.13849}, 2024.

\bibitem{sell2018diarization}
G.~Sell, D.~Snyder, A.~McCree, D.~Garcia-Romero, J.~Villalba, M.~Maciejewski, V.~Manohar, N.~Dehak, D.~Povey, S.~Watanabe, and S.~Khudanpur, ``Diarization is hard: Some experiences and lessons learned for the jhu team in the inaugural dihard challenge,'' in \emph{Proc. INTERSPEECH}, 2018, pp. 2808--2812.

\bibitem{ryant2019second}
N.~Ryant, K.~Church, C.~Cieri, A.~Cristia, J.~Du, S.~Ganapathy, and M.~Liberman, ``The second dihard diarization challenge: Dataset, task, and baselines,'' in \emph{Proc. INTERSPEECH}, 2019, pp. 978--982.

\bibitem{ryant2020third}
N.~Ryant, P.~Singh, V.~Krishnamohan, R.~Varma, K.~Church, C.~Cieri, J.~Du, S.~Ganapathy, and M.~Liberman, ``The third dihard diarization challenge,'' in \emph{Proc. INTERSPEECH}, 2021, pp. 3570--3574.

\bibitem{gao2025multimodalinformationbasedspeech}
M.~Gao, S.~Wu, H.~Chen, J.~Du, C.-H. Lee, S.~Watanabe, J.~Chen, S.~S. Marco, and O.~Scharenborg, ``The multimodal information based speech processing (misp) 2025 challenge: Audio-visual diarization and recognition,'' 2025.

\bibitem{he2016deep}
K.~He, X.~Zhang, S.~Ren, and J.~Sun, ``Deep residual learning for image recognition,'' in \emph{Proc. CVPR}, 2016.

\bibitem{vaswani2017attention}
A.~Vaswani, N.~Shazeer, N.~Parmar, J.~Uszkoreit, L.~Jones, A.~N. Gomez, L.~u. Kaiser, and I.~Polosukhin, ``Attention is all you need,'' in \emph{Proc. NeurIPS}, vol.~30, 2017.

\bibitem{gulati2020conformer}
A.~Gulati, J.~Qin, C.-C. Chiu, N.~Parmar, Y.~Zhang, J.~Yu, W.~Han, S.~Wang, Z.~Zhang, Y.~Wu, and R.~Pang, ``Conformer: Convolution-augmented transformer for speech recognition,'' in \emph{Proc. INTERSPEECH}, 2020, pp. 5036--5040.

\bibitem{deng2019arcface}
J.~Deng, J.~Guo, N.~Xue, and S.~Zafeiriou, ``Arcface: Additive angular margin loss for deep face recognition,'' in \emph{Proc. CVPR}, 2019.

\bibitem{Chung18b}
J.~S. Chung, A.~Nagrani, and A.~Zisserman, ``Voxceleb2: Deep speaker recognition,'' in \emph{Proc. INTERSPEECH}, 2018, pp. 1086--1090.

\bibitem{lin2024voxblink2}
Y.~Lin, M.~Cheng, F.~Zhang, Y.~Gao, S.~Zhang, and M.~Li, ``Voxblink2: A 100k+ speaker recognition corpus and the open-set speaker-identification benchmark,'' in \emph{Proc. INTERSPEECH}, 2024, pp. 4263--4267.

\bibitem{tang2021kespeech}
Z.~Tang, D.~Wang, Y.~Xu, J.~Sun, X.~Lei, S.~Zhao, C.~Wen, X.~Tan, C.~Xie, S.~Zhou \emph{et~al.}, ``Kespeech: An open source speech dataset of mandarin and its eight subdialects,'' in \emph{Thirty-fifth Conference on Neural Information Processing Systems Datasets and Benchmarks Track (Round 2)}, 2021.

\bibitem{zheng20233d}
S.~Zheng, L.~Cheng, Y.~Chen, H.~Wang, and Q.~Chen, ``3d-speaker: A large-scale multi-device, multi-distance, and multi-dialect corpus for speech representation disentanglement,'' \emph{arXiv preprint arXiv:2306.15354}, 2023.

\bibitem{cheng2024multi}
M.~Cheng and M.~Li, ``Multi-input multi-output target-speaker voice activity detection for unified, flexible, and robust audio-visual speaker diarization,'' \emph{arXiv preprint arXiv:2401.08052}, 2024.

\bibitem{chen2025misp}
H.~Chen, C.-H.~H. Yang, J.-C. Gu, S.~M. Siniscalchi, and J.~Du, ``{MISP-Meeting}: A real-world dataset with multimodal cues for long-form meeting transcription and summarization,'' in \emph{Proc. ACL}, 2025, pp. 1--14.

\bibitem{snyder2015musan}
D.~Snyder, G.~Chen, and D.~Povey, ``Musan: A music, speech, and noise corpus,'' \emph{arXiv preprint arXiv:1510.08484}, 2015.

\bibitem{ko2017study}
T.~Ko, V.~Peddinti, D.~Povey, M.~L. Seltzer, and S.~Khudanpur, ``A study on data augmentation of reverberant speech for robust speech recognition,'' in \emph{Proc. ICASSP}, 2017, pp. 5220--5224.

\bibitem{loshchilov2017decoupled}
I.~Loshchilov, ``Decoupled weight decay regularization,'' \emph{arXiv preprint arXiv:1711.05101}, 2017.

\bibitem{snyder2018x}
D.~Snyder, D.~Garcia-Romero, G.~Sell, D.~Povey, and S.~Khudanpur, ``X-vectors: Robust dnn embeddings for speaker recognition,'' in \emph{Proc. ICASSP}, 2018, pp. 5329--5333.

\bibitem{9414600}
J.~Thienpondt, B.~Desplanques, and K.~Demuynck, ``The idlab voxsrc-20 submission: Large margin fine-tuning and quality-aware score calibration in dnn based speaker verification,'' in \emph{Proc. ICASSP}, 2021, pp. 5814--5818.

\bibitem{Nagrani17}
A.~Nagrani, J.~S. Chung, and A.~Zisserman, ``Voxceleb: A large-scale speaker identification dataset,'' in \emph{Proc. INTERSPEECH}, 2017, pp. 2616--2620.

\end{thebibliography}

\end{document}